\documentclass[showpacs,superscriptaddress,twocolumn,amsmath,aps]{revtex4}
\usepackage{graphicx,color}
\usepackage{bm}
\usepackage[hypertex]{hyperref}

\newcommand{\be}{\begin{equation}}
\newcommand{\ee}{\end{equation}}
\newcommand{\bea}{\begin{eqnarray}}
\newcommand{\eea}{\end{eqnarray}}
\newcommand{\bsube}{\begin{subequations}}
\newcommand{\esube}{\end{subequations}}

\newcommand{\Eq}[1]{Eq.\,(\ref{#1})}

\usepackage{amsfonts}
\usepackage{amsmath}
\usepackage{amssymb}
\usepackage{graphicx}
\usepackage{times}          
\usepackage{bm}
\usepackage{color}                      

\newcommand{\alp}{\alpha}

\newcommand{\gam}{\gamma}

\newcommand{\eps}{\epsilon}
\newcommand{\vep}{\varepsilon}

\newcommand{\Gam}{\Gamma}

\newcommand{\beq}{\begin{equation}}
\newcommand{\eeq}{\end{equation}}
\newcommand{\beqn}{\begin{eqnarray}}
\newcommand{\eeqn}{\end{eqnarray}}
\newcommand{\bsub}{\begin{subequations}}
\newcommand{\esub}{\end{subequations}}

\newcommand{\ket}[1]{{\left| #1 \right\rangle }}

\newcommand{\LL}{\mathcal {L}}

\newcommand{\cdg}{c^\dagger}
\newcommand{\ddg}{d^\dagger}
\newcommand{\fdg}{f^\dagger}

\begin{document}

\title{Cross-correlations mediated by Majorana bound states}

\author{Peiyue Wang}
\affiliation{Department of Physics, Beijing Normal University,
Beijing 100875, China}

\author{Yunshan Cao }
\affiliation{School of Physics, Peking University,
Beijing 100871, China}
\affiliation{Department of Physics,
Beijing Normal University, Beijing 100875, China}

\author{Ming Gong}
\email{skylark.gong@gmail.com}
\affiliation{Department of Physics, The University of Texas at Dallas,
Richardson, Texas, 75080 USA}

\author{Gang Xiong}
\affiliation{Department of Physics, Beijing Normal University,
Beijing 100875, China}

\author{Xin-Qi Li}
\email{lixinqi@bnu.edu.cn}
\affiliation{Department of Physics,
Beijing Normal University, Beijing 100875, China}

\begin{abstract}
We consider the parallel transport through two quantum dots
correlated by a semiconductor nanowire which may support
a pair of Majorana bound states at the ends.
In addition to transient dynamics,
via modulating the quantum dot levels, we reveal a characteristic
feature of symmetry and antisymmetry in the spectral
density of cross correlation mediated by Majorana fermion.
We also find an intriguing behavior of vanished cross correlation
when one of the dot levels is in resonance with the Majorana zero mode.
\end{abstract}


\maketitle

It has attracted considerable attention in the past years
to search for Majorana fermion \cite{MF-1,MF-2} in solid states,
such as the 5/2 fractional quantum Hall system\cite{Moore1991}
and the $p$-wave superconductor and superfluid\cite{Read2000,Kit01}.
In particular, it was predicted that
a semiconductor nanowire, with strong spin-orbit coupling
and subject to external magnetic field, may support
zero-energy Majorana bound state (MBS)
when the nanowire is in proximity to
an $s$-wave superconductor\cite{Sau10,Sau12, Ore10}.
This proposal has the major advantage of requiring only the most
conventional materials, being thus easier to implement in experiments.
Indeed, in some recent experiments\cite{Kou12,MF2,MF3,MF4}, evidences
of the MBSs were spotted in this sort of hybrid nanowire systems.

The realization and manipulation of Majorana fermion in solid-states
may pave a way for the desirable fault-tolerant
topological quantum computation\cite{Kit03,Ste10,Nay08,Ali11}.
In pure physics, the Majorana study in solid states
is intriguing because of many unusual effects,
such as the non-Abelian statistics \cite{Alicea11},
the sharp jump in conductance peak \cite{Liu11},
the peculiar noise behaviors \cite{Bol07,Law09,Li12,Shen12,Zoch12},
and the 4$\pi$ periodic Majorana-Josephson currents
\cite{Kit01,Lut10,Ore10,Fu09}, etc.
Remarkably, most of these effects are associated with
the {\it nonlocal} nature of the Majorana fermion\cite{Nil08,LeeDH08,Li12-a}.
The novel nonlocality feature can be most surprisingly
elaborated by the example of its nanowire realization.
In this case, the emerged pair of MBSs
at the ends of the nanowire constitutes, respectively,
the {\it real} and {\it imaginary} parts of an ordinary fermion.
This means that, if an electron with energy smaller than the energy gap
between the Majorana zero mode and other exited states is injected into
the nanowire, the electron would split into
two Majorana bound states which are essentially correlated
but spatially separated.
In some sense the Majorana's nonlocality can be related to a nonlocal
cross Andreev reflection (CAR) process \cite{Law09,Nil08},
because of the presence of superconductivity.
However, as described above, the concept of nonlocality
of Majorana fermion itself has richer information
and is more intriguing than the CAR phenomenon.
In Ref.\ \cite{Sau04} the nonlocal CAR process
has been exploited for quantum teleportation,
following the same idea of the pioneering work
by Bennett {\it et al.} \cite{Ben93}.
In such teleportation scheme, what is teleportated
is the ``state information", but not a ``matter"
as possibly mediated by the Majorana fermion, based on its
nonlocality nature \cite{LeeDH08,Li12-a,Fu10}.   

In this work, we attempt to study the Majorana's nonlocality
in a rather direct way. As schematically shown in Fig.\ 1,
we consider a setup of parallel transports
through two quantum dots (QDs) coupled
by a semiconductor nanowire which may
support a pair of MBSs at the ends.
We will focus on the cross correlation of currents
mediated by the MBSs and reveal some unusual behaviors.   

\begin{figure}
 \center
 \includegraphics[scale=0.5]{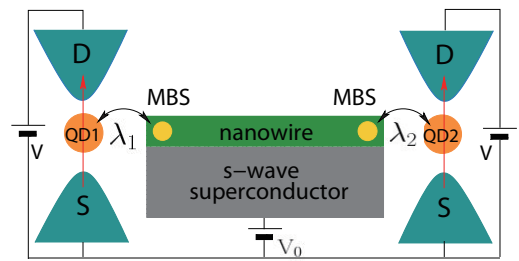}
 \caption{\label{4}
(Color online)
Schematic setup for a cross-correlated parallel transport
through two quantum dots which are weakly tunnel-coupled
by a semiconductor nanowire.
The nanowire is in contact with an $s$-wave superconductor.
Under appropriate conditions, such as with large Zeeman splitting
and strong spin-orbit interaction, a pair of Majorana bound states
is anticipated to emerge at the ends of the nanowire.
Here we explicitly display the closed circuit that satisfies
the law of current conservation. }
\end{figure}

\vspace{0.2cm}
{\it Model and Method}.---
The transport setup is schematically shown in Fig.\ 1,
where the semiconductor nanowire is in proximity to
a conventional $s$-wave superconductor
and subject to a Zeeman field.
When the Zeeman splitting energy $V_z$,
the proximity-induced order parameter $\Delta$,
and the chemical potential $\mu$
satisfy the condition $V_z > \sqrt{\Delta^2 + \mu^2}$,
the nanowire is driven into a topological
superconducting phase and a pair of zero-energy MBSs
will emerge at the ends of the nanowire \cite{Ore10, Sau10,Sau12}.
As a result, the two quantum dots which are tunnel-coupled to the
ends of the nanowire are now {\it effectively} coupled to the MBSs.
The whole system is described by
\begin{equation}
		H=H_{\rm leads}+H_T + H_{\rm sys}.
\end{equation}
$H_{\rm leads}=\sum_{j=1,2}\sum_{\alpha=S,D}
\sum_k\varepsilon_{kj\alpha} c_{kj\alpha}^\dagger c_{kj\alpha}$,
is the Hamiltonian of the source (S) and drain (D) leads
for transports through the left ($j=1$) and right ($j=2$) dots.
$H_T=\sum_{j=1,2}\sum_{\alpha=S,D}
\sum_k t_{j\alpha}d_j^\dagger c_{kj\alpha}+{\rm H.c.}$
describes the tunneling between the leads and the dots.
The last term is the summed {\it system} Hamiltonian
of the quantum dots and the tunnel-coupled MBSs
\cite{Liu11,Bol07,Law09,Lut10,LeeDH08}
\begin{eqnarray}
\label{M-QD-Hami}
H_{\rm sys} &=& \sum_{j=1,2}\epsilon_j d_j^\dagger d_j
+ \frac{i}{2}\epsilon_M\gamma_1\gamma_2   \nonumber\\
			  &+&  \lambda_1(d_1^\dagger-d_1)\gamma_1
+ i\lambda_2(d_2^\dagger+d_2)\gamma_2 .
\end{eqnarray}
In the above Hamiltonians,
$\ddg_j(d_j)$ and $\cdg_{j\alp k}(c_{j\alp k})$ are,
respectively, the electron creation (annihilation) operators
of the quantum dots and the attached transport leads,
with energies $\eps_j$ and $\vep_{kj\alp}$.
$\gamma_1$ and $\gamma_2$ are the Majorana operators, while
$\epsilon_M=\Delta e^{-\ell/\xi}$ is the overlap coupling amplitude
between the MBSs, with $\ell$ the wire length
and $\xi$ the superconducting coherence length.

For the sake of calculation convenience, we move from the
Majorana to regular fermion ($f$ and $\fdg$) representation,
via the transformation of $\gam_1=\fdg+f$ and $\gam_2=i(\fdg-f)$.
The system Hamiltonian is accordingly rewritten as
\begin{eqnarray}
H_{\rm sys} &=& \epsilon_M f^\dagger f + \sum_{j=1,2} \big\{
             \epsilon_j d_j^\dagger d_j \nonumber \\
            && + \lambda_j [(d_j^\dagger + \eta d_j)f + {\rm H.c} ] \big\}.
\label{M-QD-H}
\end{eqnarray}
Here we introduce a parameter $\eta$ to distinguish
the Majorana fermion ($\eta = 1$) from the regular
bound state (RBS, $\eta = 0$).
This will be useful for latter comparison.
Associated with the above {\it system} Hamiltonian, we introduce the state
basis as $\ket{n_1n_fn_2}$, where $n_{1,2,f}=0$ or $1$ stand for
the occupation number of the QDs and the MBSs.
One may notice that the tunneling terms in $H_{\rm sys}$
only conserve charge modulo $2e$. This reflects the
fact that a pair of electrons can be extracted out from the
superconductor and can be also absorbed by the condensate.
As a consequence, for the MBSs, the Hilbert space of
the central system can be split into two subspaces:
$\{\ket{100},\ket{010},\ket{001},\ket{111}\}$ with odd parity;
and $\{\ket{011},\ket{101},\ket{110},\ket{000}\}$ with even parity.
For the RBS, the Hilbert space is the same
but does not need to be split by parity consideration.
In the following these basis states will be used to calculate
the transport currents and their correlation,
based on a master equation approach.   


For quantum transport, the leads can be regarded as a general
environment, while the central device is the {\it system
of interest} described by a reduced density matrix $\rho(t)$.
Moreover, in weak transmission limit, the Born-Markov
approximation applies well, for the setup of Fig.\ 1
which would result in the following master equation \cite{Mil99}
\begin{equation}\label{u-ME}
\dot{\rho} = -i\mathcal{L}\rho +\sum_{j = 1,2}\big(
\Gamma_{j}^S\mathcal{D}[d_j^\dagger]\rho
 +\Gamma_{j}^D\mathcal{D}[d_j]\rho\big) ,
\end{equation}
where $\LL \rho\equiv [H_{\rm sys},\rho]$ and
${\cal D}[A]\rho\equiv A\rho A^{\dagger}-\frac{1}{2}\{A^{\dagger}A, \rho\}$.
In \Eq{u-ME} we employ the tunneling rates $\Gam_{j}^{S(D)}$
for the $j$th dot coupling to the source ($S$) and drain ($D$) leads,
respectively. These rates, more specifically, can be determined
by the tunnel-coupling amplitudes and the density-of-states of the leads. 

The above Born-Markov master equation is valid
in the weak transmission limit.
This is also equivalent to a large bias limit,
since both correspond to ignoring the level's broadening effect,
under the condition that the dot level is far from
the Fermi levels of the leads, with at least several $\Gamma$
($\Gamma$ stands for the level's broadening).
The large bias condition has been examined carefully in Ref.\ \cite{LiF09},
while in \cite{LiJ11} an improved scheme is proposed
to make the new master equation applicable to arbitrary voltages.
Therefore, for the setup under study, \Eq{u-ME} is valid
in the large bias regime.
Similar equation under the Born-Markov approximation
has been applied in the recent work
for Majorana systems as well \cite{Shen12,Flen11}.
The {\it large bias} condition may raise
a concern for its effect on the Majorana bound states.
In the context of topological quantum computation,
it was analyzed that a tunnel-coupling to electronic reservoir
would be detrimental to the Majorana qubit \cite{Bud12}.
However, in our present work
(as well as in any other Majorana detection scheme),
the major concern is not the Majorana qubit,
but the Majorana signature in transport currents.
In principle, from the general rule of measurement,
the backaction (decoherence) is unavoidable.
However, it does not destroy the Majorana signature.
In either the steady-state current or noise spectrum,
the transport backaction only results in a broadening
effect on the conductance or noise spectral peak.
Finally, it should be noted that a large bias does not necessarily
imply a strong transport current, since the coupling
between the dot and leads can be very weak.  

Based on \Eq{u-ME}, the source and drain currents
through each dot can be simply calculated by
\begin{equation}
I_{j}^{S}= e \Gam^S_j (1-n_{j}),
\quad I_{j}^{D} = e \Gam^D_j n_{j},
\label{eq-Is}
\end{equation}
where the occupation of the corresponding quantum dot
is calculated by $n_j=\mathrm{Tr}[d_j^\dagger d_j\rho(t)]$.
In particular, using the stationary density matrix ($\bar{\rho}$),
the steady-state currents can be obtained.

The master equation also allows us to calculate,
very conveniently, the correlation of currents \cite{Mil99}.
In this work, we focus on the cross correlation
of currents through the different dots:
$G_{ij}(t_2,t_1)=\mathrm{E}[I_i(t_2)I_j(t_1)]$,
where $E[\cdots]$ stands for the ensemble average
in the quantum trajectory approach\cite{Mil99}.
More specifically, we consider only the cross correlation
of the drain currents (denoted here by $I_{j(i)}$ for brevity).
The current is defined through $I_j(t)= e ~ dN_j(t)/dt$
and the average is determined by
$\mathrm{E}[dN_j(t)]=dt\mathrm{Tr}[d_j^\dagger d_j\rho(t)]$.
In steady state, the cross correlation is a function of the
time difference owing to the time-translational invariance,
i.e., $ G_{ij}(t_2,t_1)=G_{ij}(t_2-t_1)\equiv G_{ij}(t)$.
Therefore we have
\begin{eqnarray}
G_{ij}(t)
 &=& e^2\mathrm{Tr}\{\Gamma_{i}^D d_i[e^{\widetilde{\mathcal{L}}t}
  (\Gamma_{j}^D d_j \bar{\rho}d_j^\dagger)]d_i^\dagger\}         \nonumber\\
&&  =e^2 \Gamma_{i}^D\Gamma_{j}^D \mathrm{Tr}[d_i \tilde{\rho}_j(t)d_i^\dagger] .
\end{eqnarray}
Here the Liouvillian propagator $e^{\widetilde{\mathcal{L}}t}$
describes the state evolution governed by \Eq{u-ME}.
From the post-selected state after a jump, i.e.,
$\tilde{\rho}_j(0)=d_j\bar{\rho}d_j^\dagger$, we introduced
$\tilde{\rho}_j(t)\equiv e^{\widetilde{\mathcal{L}}t}\tilde{\rho}_j(0)$
which obviously satisfies \Eq{u-ME}.

The spectral density of cross correlation is given by
the Fourier transform of $G_{ij}(t)$, which yields
\begin{eqnarray}
S_{ij}(\omega)
= e^2 \Gamma_{i}^D\Gamma_{j}^D \mathrm{Tr}
 [ d_i \tilde{\rho}_j(\omega)d_i^\dagger
 +  d_j \tilde{\rho}_i(-\omega)d_j^\dagger ].
\end{eqnarray}
In deriving this result, we have used the property
$G_{21}(t)=G_{12}(-t)$ and introduced the definition
$\tilde{\rho}_j(\pm\omega)=\int_0^\infty\tilde{\rho}_j(t)
e^{\pm i\omega t}dt$.   
Straightforwardly, via the technique of Laplace transformation,
$\tilde{\rho}_j(\omega)$ can be solved from \Eq{u-ME}.
Moreover, we would like to employ a symmetrized
spectral density, $S(\omega)=[S_{21}(\omega)+S_{12}(\omega)]/2$,
to characterize the cross correlation. We have
\begin{eqnarray}\label{SW}
S(\omega)
 =e^2\Gamma_{1}^D\Gamma_{2}^D\mathrm{Re}
\{ \mathrm{Tr}[ d_1\tilde{\rho}_2(\omega)d_1^\dagger
+  d_2 \tilde{\rho}_1(\omega)d_2^\dagger] \}.
\end{eqnarray}
Also, we will use the Fano factor to describe
the zero-frequency correlation.
The Fano factor is defined as $F=S(0)/2e\sqrt{\bar{I}_1 \bar{I}_2}$,
where $\bar{I}_{1(2)} = [\bar{I}_{1(2)}^S + \bar{I}_{1(2)}^D]/2$
is the average stationary current through the left (right) dot.  

In Refs.\ \cite{LeeDH08,Li12-a}, the isolated {\it dot-MBSs-dot} system
was studied. Remarkably, it was found that, even in the limit $\epsilon_M=0$,
a novel process of {\it teleportationlike} electron transfer
between the distant dots is possible,
due to the inherent {\it nonlocal} nature of the Majorana fermion.
If $\epsilon_M\neq 0$, it also leads to a {\it discontinuity}
of the tunneling conductance through one of the dots,
under the MBS-mediated influence of the other dot.
In the following, we will show:
{\it (i)} In the limit $\epsilon_M=0$, there is no cross-correlation
between the transport currents through the two individual dots.
This is an interesting result, by noting
the presence of electron transfer between the dots, mediated by the MBSs.
{\it (ii)} For $\epsilon_M\neq 0$, distinct Majorana signatures
can appear in the cross correlation of currents.

\begin{figure}
  \center
  \includegraphics[scale=0.36]{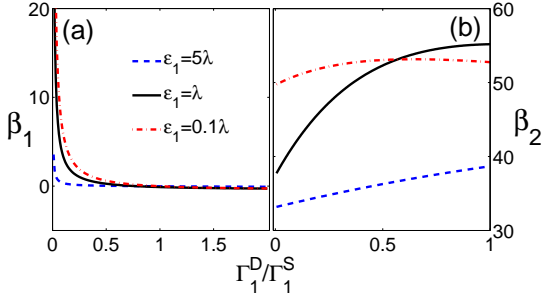}
  \caption{\label{FIG2}
  (Color online) Cross-correlation of the steady state currents,
  plotted by the scaled difference of the source and drain currents,
  $\beta_j = (\bar{I}_{j}^S-\bar{I}_{j}^D)/\bar{I}_{j}^D$,
  with $j=1(2)$ denoting the left (right) dot.
  The cross-correlation is manifested by modulating
  $\Gamma^D_1/\Gamma^S_1$ and $\epsilon_1$ of the left dot,
  via their effects on $\beta_2$ of the right dot.
Parameters:
$\lambda_1=\lambda_2=\lambda$, $\epsilon_M=\epsilon_2=\lambda$,
$\Gamma_1^S =\Gamma_2^S=\lambda$, and $\Gamma_2^D=0.01\lambda$.    }
\end{figure}

\vspace{0.2cm}
{\it Stationary and Transient Correlation}.---
Let us consider first the steady-state currents through the two dots
and their correlation mediated by the MBSs.
We notice that, owing to coupling to the MBS,
in the transport through a quantum dot the source current
is in general {\it not} equal to the drain current \cite{Li12}.
Their difference, in essence, is compensated by a current flowing
through the nanowire into or from the superconductor condensate.
For our present double-dot system, owing to the similar reason,
we introduce $\beta_j = (\bar{I}_{j}^S-\bar{I}_{j}^D)/\bar{I}_{j}^D$ ($j=1,2$)
to qualify this unique feature.
In Fig.\ 2, we use this factor to illustrate the cross correlation.
Specifically, we alter $\Gamma^D_1/\Gamma^S_1$,
and examine its effect on $\beta_1$ and in particular on $\beta_2$.
The results in Fig.\ 2 correspond to $\epsilon_M\neq0$,
where the existence of cross correlation is obviously indicated,
while its peculiar feature is to be further exploited in the following.
We would like to mention that, if $\epsilon_M=0$,
altering $\Gamma^D_1/\Gamma^S_1$ would have no effect on $\beta_2$.
This result completely differs from what happens for two quantum dots
coupled through a regular bound state.

\begin{figure}
 \center
 \includegraphics[scale=0.35]{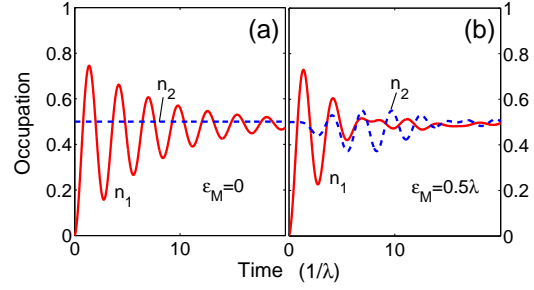}
 \caption{ (Color online)
 Transient dynamics of the right dot in response to
 an electron jump from the left dot to its drain lead,
 based on the steady state.
 The occupation of the $j$th dot is obtained by
 $n_j(t)={\rm Tr}[d_j^\dagger d_j \rho(t)]$,
 with $\rho(t)$ evolved from the specific ``initial" state
$\rho(0)=d_1\bar{\rho}d_1^{\dagger}/{\rm Tr}(d_1\bar{\rho}d_1^{\dagger})$.
Assuming $\lambda_1=\lambda_2=\lambda$,
we give the result in (a) for $\epsilon_M=0$
and in (b) for $\epsilon_M=0.5\lambda$.
Other parameters: $\epsilon_1=\epsilon_2=\lambda$,
and $\Gamma_{S(D)}^{1(2)}=0.1\lambda$.  }
\end{figure}

To further illustrate the MBS-mediated cross correlation,
we now turn to the {\it transient} dynamics.
To be specific, we consider how a {\it disturbance}
on the left dot would affect the transport through the right one.
Based on the steady state, the disturbance can be introduced by a
{\it tunneling event} of an electron from the left dot to its drain lead.
This consideration defines a particular ``initial" state,
$d_1\bar{\rho}d_1^{\dagger}/{\rm Tr}(d_1\bar{\rho}d_1^{\dagger})$.
Then, we consider the transient currents associated with this new state.

First, in Fig.\ 3(a) we show the result in the limit of $\epsilon_M=0$.
We find that the right current is unaffected by the disturbance on
the left dot, which is in agreement with the above tunnel-rate modulation.
This result straightforwardly indicates a zero cross-correlation
between the currents through the two dots.
This behavior is indeed not obvious at all, by two observations:
{\it (i)}
If we consider the two dots being coupled
by a regular bound state, even for non-aligned levels of
$\epsilon_1=\epsilon_2=\lambda$ and $\epsilon_M=0$, we find
that the cross-correlation between $I_1$ and $I_2$ is nonzero.
{\it (ii)}
It was found in Refs.\ \cite{LeeDH08,Li12-a},
where the isolated dot-MBSs-dot system was analyzed,
that the state $\ket{100}$ can evolve to $\ket{001}$
in short timescale even in the limit $\epsilon_M=0$.
This seemingly indicates that the right current $I_2$ should
be affected by the left-dot disturbance in this limit.

In Fig.\ 3(b) we show the result of $\epsilon_M\neq 0$, assuming
$\epsilon_M=0.5\lambda$ and $\epsilon_1=\epsilon_2=\lambda$.
In the transformed basis, this setup
does not look too much different from $\epsilon_M=0$.
However, we find in Fig.\ 3(b) that the current through the right dot
is considerably affected by the disturbance on the left dot.
Moreover, as to be shown in the following, its correlation behavior
differs significantly from that mediated by a regular bound state
with this same energy ($\epsilon_M$).

\begin{figure}
 \center
 \includegraphics[scale=0.5]{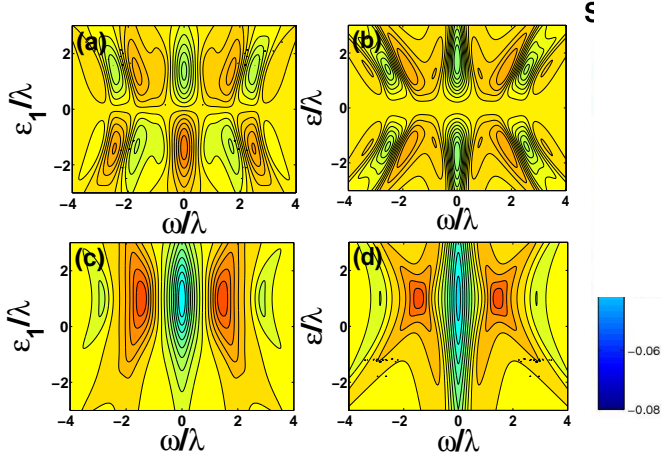}
 \caption{ (Color online)
Spectral density of the cross correlation {\it versus}
two types of dot-level modulations:
in (a) and (c) modulation of the left dot level ($\epsilon_1$) only;
and in (b) and (d) simultaneous modulation of the both dot levels
($\epsilon_1=\epsilon_2=\epsilon$).
For the single dot $\epsilon_1$-modulation,
we assume $\epsilon_2=\epsilon_M=\lambda$,
where $\lambda$ is the common coupling strength between
the QD and the MBS (i.e., $\lambda_1=\lambda_2=\lambda$).
In (a) and (b) we show the results for the MBSs with
$\epsilon_M=\lambda$, while the results in (c) and (d)
correspond to a regular bound state (RBS) with energy
$\epsilon_M$, as a comparison.
In all these results, we have set $\Gamma^{S(D)}_{1(2)}=0.3\lambda$.   }
\end{figure}

\vspace{0.2cm}
{\it Cross Correlation Spectrum}.---
Below we employ the spectral density function Eq.\ (8)
to analyze the transient dynamics.
In Fig.\ 4 we present a contour-plot for the spectral density of the
cross correlation, as a function of frequency and dot-level modulations.
In Fig.\ 4(a) and (b) we show the results for MBSs
and in (c) and (d) for RBS, using the same energy
($\epsilon_M\neq 0$) for the sake of comparison. 
We find all the results in Fig.\ 4(a)-(d)
an even function of the frequency. This is because of the property
$G_{ij}(t)=G_{ji}(-t)$ and the symmetrization procedure
for the cross-correlation function which leads to \Eq{SW}.
More interesting is that, for the MBS-mediated correlation,
the spectral density function would have different
dependence on the dot-level modulations.
That is, for the level modulation of single dot,
the spectral density in Fig.\ 4(a) is {\it antisymmetric}
with respect to the energy of the Majorana zero mode.
In contrast, however, the spectrum in Fig.\ 4(b) is {\it symmetric}
under the simultaneous modulation of the double-dot energy levels.  
We conclude that this is one of the important signatures
resulted from the Majorana nonlocal correlation.
In Fig.\ 4(c) and (d) we observe that the cross-correlation
mediated by a regular bound state, differently, are always
{\it symmetric} under both sorts of dot-level modulations.

\begin{figure}
 \center
 \includegraphics[scale=0.37]{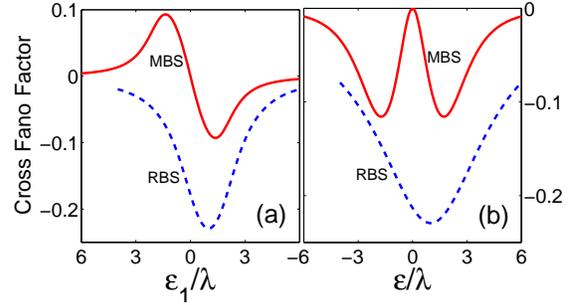}
 \caption{(Color online)
Fano factor of the cross correlation, which corresponds to
the zero-frequency noise and is defined in the main text.
All the parameters used in this figure are the same as in Fig.\ 4. }
\end{figure}

To show more clearly the symmetry behavior mentioned above,
we plot in Fig.\ 5 the zero-frequency Fano factor.
We find that the results for MBSs, being either symmetric
or antisymmetric, are centered at the Majorana zero mode
and have a {\it vanished} cross Fano factor there.
This result cannot be interpreted by the usual picture
of resonant transfer in terms of matching the energy levels.
Actually, in the transformed representation,
the two MBSs are ``combined" to form
a regular fermion with energy $\epsilon_M$.
This is likely to indicate that the cross correlation,
including the frequency-dependent spectrum in Fig.\ 4(a) and (b),
should be centered at the ``effective" energy $\epsilon_M$.
Moreover, when one of the dot levels is
in resonance with the Majorana zero mode,
the {\it vanished} cross correlation is far beyond our intuition. 
As a sharp comparison, in Fig.\ 5
we display the results from the RBS-mediated correlation.
Evidently, we find that the cross Fano factor is always
symmetric with respect to the {\it RBS energy},
for the both types of dot-level modulations.
Moreover, at the symmetry center
the cross correlation is {\it nonzero}.
Therefore, these remarkable differences,
together also with the entire
``lineshape" in response to the dot-level modulations,
provide us very useful information
to identify the emerged Majorana bound states.   

We became aware of a couple of recent studies on
the Majorana-mediated cross correlation.
In Ref.\ \cite{Bru11} the motion (transport) of Majorana fermion
through a Hanbury Brown-Twiss type interferometer is considered.
In this four terminal transport setup, it is found that the cross correlation
of currents reveals a novel ``+(-)" sign signature and an absence of
partition noise after adding a quantum-point-contact,
both of which can be traced back to the Majorana nature of the carriers.
Being more relevant to the present work,
in Refs.\ \cite{Shen12,Zoch12},
the series transport through two quantum dots coupled
by a pair of MBSs is analyzed,
where the analogous suppressed
cross correlation is observed as well,
under the same condition as mentioned above.
Therefore, the vanished cross correlation found in both
the parallel and series transport setups
should originate from the same Majorana physics,
being of interest and worth further clarification.    

\vspace{0.2cm}
{\it Experimental Feasibility}.---
For possible experimental demonstration, the semiconductor nanowire
can be the InSb wire, as utilized in the recent experiment \cite{Kou12}.
The InSb nanowire has a large $g$-factor ($g\simeq 50$),
and a strong Rashba
spin-orbit interaction (with energy $\sim$50 $\mu$eV).
For magnetic field of $0.15$ Tesla,
the Zeeman splitting ($V_z\simeq 225~\mu{\rm eV}$)
starts to exceed the induced superconducting gap
$\Delta\simeq 200~\mu{\rm eV}$.
Moreover, a low temperature such as $T=100$ mK
can suppress the thermal excitation of the Majorana zero mode
to higher energy states. Under these conditions,
the MBSs may appear at the ends of the nanowire \cite{Kou12}.
In practice, the energy level modulation of the quantum dot
can be implemented by gate voltage.
Utilizing the gate-voltage technique, one can also control the
tunneling rates $\Gamma^{1(2)}_{S(D)}$ and the coupling energy $\lambda$
to the order of a few $\mu$eV, which can maintain electron
oscillations between the QD and MBSs
within a coherence time longer than nanoseconds.
This magnitude order of coupling values also allows
the condition $k_BT<eV<E_z$ to define
a broad range of bias voltage across the quantum dot,
satisfying the {\it large} bias
condition to guarantee the Born-Markov master equation.

\vspace{0.2cm}
{\it Summary.}---
We analyzed the cross correlation mediated by a pair of Majorana
bound states, based on a specific setup
of parallel transport through two quantum dots.
The essential ``dot-MBSs-dot" building block of the setup
is experimentally accessible
using the nowaday state-of-the-art technology,
particularly based on a semiconductor nanowire realization.
From the cross correlation of the parallel currents,
we revealed several unusual Majorana signatures,
including the symmetry and antisymmetry behaviors
of the spectral density in response to the dot-level modulations,
and the vanished cross correlation when any
of the dot-levels is in resonance with the Majorana zero mode.
These signatures are of interest and beyond possible interpretations
based on the conventional regular fermion system.   

\vspace{0.5cm}
{\it Acknowledgments.}---
This work was supported by the Major State Basic
Research Project (Nos.\ 2011CB808502 \& 2012CB932704),
and the NNSF of China (No.\ 101202101).

\end{document}